\RequirePackage{fix-cm}

\documentclass[]{svjour3}[]

\smartqed

\usepackage[utf8x]{inputenc}
\usepackage{graphicx}
\usepackage{array}
\usepackage{amsfonts}
\usepackage{amsmath}
\usepackage{amssymb}
\usepackage{booktabs}
\usepackage{bold-extra}
\usepackage{float} 
\usepackage{bm} 
\usepackage{ifthen}
\usepackage[toc,page]{appendix} 
\usepackage{stmaryrd}

\usepackage{aas_macros}  

\usepackage[round, authoryear]{natbib}
\newcolumntype{L}{>{$}l<{$}} 

\DeclareFontFamily{U}{wncy}{}
\DeclareFontShape{U}{wncy}{m}{n}{<->wncyr10}{}
\DeclareSymbolFont{mcy}{U}{wncy}{m}{n}
\DeclareMathSymbol{\diracComb}{\mathord}{mcy}{"58} 

\title{
Building the analytical response in frequency domain of AC biased bolometers }

\subtitle{Application to {\em Planck/HFI} }

\author{
	Alexandre Sauv\'e 
	\and
	Ludovic Montier
	}

\numberwithin{equation}{section} 

 \journalname{Experimental Astronomy}

\begin{document}

	\institute{A. Sauv\'e \at
               	IRAP (CNRS)\\
		9, avenue du Colonel Roche - BP 44346 - 31028 Toulouse Cedex 4 - FRANCE\\
		Tel : +33 5 61 55 66 66 \\
		Fax : +33 5 61 55 86 92 \\
              	\email{asauve@gmail.com}           
           \and
           	L. Montier \at
              	IRAP (CNRS) \\
		\email{ludovic.montier@irap.omp.eu}
	}

\newcommand{\Planck}		{{\em Planck}} 
\newcommand{\HFI}			{{\em Planck/HFI}} 

	\maketitle
\Planck{}

	\begin{abstract}

	{\textsc{Context}:} Bolometers are high sensitivity detector commonly used in Infrared astronomy. 
	The {\em HFI} instrument of the \Planck{} satellite makes extensive use of them, 
	but after the satellite launch two electronic related problems revealed critical.
	First an unexpected excess response of detectors at low optical excitation frequency for $\nu < 1$ Hz,
	and secondly the Analog To digital Converter (ADC) component had been insufficiently characterized on-ground.
	These two problems require an exquisite knowledge of detector response.
	However  bolometers have highly nonlinear characteristics, coming from their electrical and thermal coupling 
	making them very difficult to modelize.

	{\textsc{Goal}:} We present a method to build the analytical transfer function in frequency domain which describe the voltage response
	of an Alternative Current (AC) biased bolometer to optical excitation, based on the standard bolometer model. 
	This model is built using the setup of the \HFI{} instrument and offers the major improvement of being based on
	a physical model rather than the currently in use had-hoc model based on Direct Current (DC) bolometer theory.

	{\textsc {Method}:} The analytical transfer function expression will be presented in matrix form. For this purpose,
	we build linearized versions of the bolometer electro thermal equilibrium. And a custom description
	of signals in frequency is used to solve the problem with linear algebra. 
	The model performances is validated using time domain simulations.

	{\textsc{Results}:} The provided expression is suitable for calibration and data processing. 
	It can also be used to provide constraints for fitting optical transfer function using real data from 
	steady state electronic response and optical response. 
	The accurate description of electronic response can also be used to improve the ADC nonlinearity
	correction for quickly varying optical signals.

	\end{abstract}




\tableofcontents

\newcommand{\Wm}			{\ensuremath{\Omega_m}}
\newcommand{\Fm}			{\ensuremath{f_{m}}}
\newcommand{ \dt}[1]    {  \ensuremath{ \dot{ #1}}  }
\newcommand{ \order}    { \ensuremath{ \mathcal O } }
\newcommand{ \Vavg } {\ensuremath{\langle V \rangle}}
\newcommand{ \MV }     { \ensuremath{ \overline{V}} }
\newcommand{ \EV }     { \ensuremath{ \widetilde{V} }}
\newcommand{ \MT }     { \ensuremath{ \overline{T}} }
\newcommand{ \ET }     { \ensuremath{ \widetilde{T} }}
\newcommand{ \MR }     { \ensuremath{ \overline{R} }}
\newcommand{ \ER }     { \ensuremath{ \widetilde{R} }}
\newcommand{ \Rref }     { \ensuremath{ \langle \MR \rangle } }
\newcommand{ \Tref }     { \ensuremath{ T_e} }
\newcommand{ \Ravg } {\ensuremath{\langle R \rangle}}
\newcommand{\Tavg}[1][1]	{\ensuremath{\langle T \ifthenelse{1=#1}{}{_{#1}} \rangle}}
\newcommand{ \Popt } {\ensuremath{ P_{opt} }}
\newcommand{ \Pavg } {\ensuremath{\langle \Popt \rangle}}
\newcommand{ \PR }     { \ensuremath{ \widetilde{P}_{opt} }}
\newcommand{ \dVbias }     { \ensuremath{ \dot{V}_{bias}} }
\newcommand{ \Rsteady }  { \ensuremath{(\Ravg + \MR)} }

\section{Introduction}
\label{introduction}

Bolometers are high sensitivity thermal detectors commonly used in astronomy 
in the domain of infrared to sub-millimeter wavelengths.
They are basically semi conductor thermometers, connected to a heat sink, which impedance vary with temperature.
AC biased bolometers have been extensively used for the last ten years in 
balloon borne and space experiments as in the \Planck{} satellite, 
mainly for their good performances in regard to low frequency $1/f$ noise. 
However AC biased bolometers detectors are still described based on DC theory by \citet{2008ApOpt..47.5996H}.
Some work has been done by \citet{2010ApOpt..49.5938C} for optimizing the AC biasing of bolometers
in the case of the \HFI{} instrument, but without describing the shape of the electronic response.

The \Planck{}  experiment \citep{2010A&A...520A...1T}, designed to observe the Cosmic Microwave Backgroud (CMB),
reached an unprecedented sensitivity with $\frac{\Delta T}{T}$ better than $10^{-5}$ for
the CMB anisotropies observation. With respect to this objective, two problems related to electronics
have revealed as critical after the satellite launch: the low frequency excess response (LFER) and
Analog to Digital Converter (ADC) nonlinearity.

First the detectors response to optical excitation exhibited an excess response at low frequencies
for $\nu < 1 Hz$ \citep{2015arXiv150201586P}. The main culprit for this excess is intermediate components in the
thermal path to the heat sink operating at $100$ mK. These components produce also specific response to energy
deposit from particles \citep{2014A&A...569A..88C}. The thermal model have been extended
with a chain of order 1 low pass filters to build an ad-hoc transfer function model \cite{2014A&A...571A...7P} which described well 
the detector response at first order. However the last version of the model needs to fit up to seven  thermal
components \citep{2015arXiv150201586P}.

The second in flight issue is the ADC nonlinearity which has been insufficiently characterized on ground. 
This systematic effect becomes very difficult to correct in the \HFI{} case, because signal is averaged onboard
over 40 samples of the modulation half period, before being sent to the ground. A very good knowledge of time domain signal at 40 times
the modulation frequency is then required to apply the ADC nonlinearity correction, 
Currently an empirical model based on the hypothesis of slowly varying signal is in use by \citet{2015arXiv150201586P},
with limited performances in the case of bright and quickly varying signal.

In order to address these very demanding objectives, the present article describes, 
an analytical model built in frequency domain for the voltage response to an 
optical excitation for an AC biased bolometer. 
This model is based on the physical model of the bolometer and has a selectable bandwidth limit.
The only free parameter is the optical excitation angular frequency $\omega$.

We will first describe in Sect.~\ref{sect_bolo_model}
the electrical and thermal model of the \HFI{} bolometer detectors.
Then in Sect.~\ref{sect_linear_versions} we show how to build a suitable linearized version of the electro-thermal equilibrium equations.
The solving of the equilibrium will be done in frequency domain and involve convolution
of frequency vectors. 
To make the solving possible, In Sect.~\ref{section_formal} a custom frequency representation of signals 
and matrix formalism is presented.
Finally in Sect.~\ref{sect_results} the model performances will be compared to the time domain simulations computed
with the Simulation for the Electronic of Bolometers ({\tt SEB}) tool, used in the \Planck{} consortium
since 2007.

\section{Bolometer model}
\label{sect_bolo_model}

In this section, we will describe the \HFI{} detector model, which involves the electronic design and the thermal model. 
The NTD-Ge  semi-conductor bolometers used have a negative thermal response. 
Which means their impedance decrease when their temperature increase. So when temperature raise from incoming
optical radiative power  the dissipated heat from Joule effect decrease. 
In this case, the {\em electro-thermal coupling} helps the system 
to reach a stable equilibrium.
The coupled electronic and thermal equations will be used in Sect.~\ref{sect_linear_versions} to build the linear response.

\subsection{Bias circuit}
\label{sect_bias}

\begin{figure}[t!]
    \centering
    \includegraphics[width=0.60\textwidth]{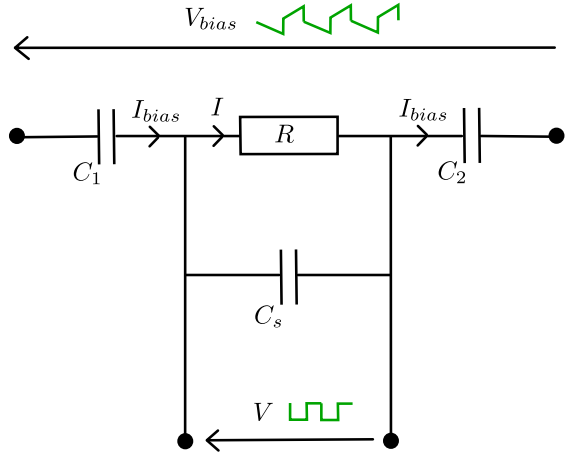}
    \caption{Bolometer bias circuit of \HFI{}. The schematic is reduced to the minimal set involved in the description of electro-thermal coupling. 
    $R$ is the bolometer impedance.
    $C_1$ and $C_2$ are the bias capacitances, doing the derivation operation on the bias voltage. 
    $C_s$ is the stray capacitance coming from cables length.
    $V_{\rm bias}$ is the circuit bias voltage composed of a triangle wave plus a square wave added to compensate for $C_s$. 
    $I_{bias}$ is the bias current provided to the $R+C_s$ couple.
    $I$ is the current flowing through the bolometer, it is designed to be as constant in absolute value as possible, 
    	to mimic a $DC$ bias. 
    And $V$ is the output bolometer voltage.
    }
    \label{bias_circuit}
\end{figure}

The \HFI{} bias circuit is presented in Fig. \ref{bias_circuit} which is an excerpt from the readout chain described in 
\Planck{} Pre-Lauch paper \citep[Sect.~4]{2010A&A...520A...9L} and in \citet{montier:thesis}.

The differential equation describing the voltage of the bolometer is
\begin{equation}
	 V(t) + R(t) ( C_{eq} + C_s ) \frac{dV(t)}{dt}  = R(t) C_{eq} \frac{d V_{\rm bias}(t)}{dt}\,,
	 \label{electrical_equilibrium} 
\end{equation}
where $V_{\rm bias}$ is the input bias voltage, $V(t)$ is the bolometer measured output voltage, 
$R(t)$ is the bolometer real impedance,  $C_1$ and $C_2$ are the bias capacitances,
with $C_{eq} = C_1 C_2 / (C_1+C_2)$, and $C_s$ is a stray capacitance. 
All quantitative values are referenced in Table~$\ref{table_elec_params}$.

Using values from Table~\ref{table_elec_params}, $C_s \simeq 60 C_{eq}$.
The very high value of  $C_s$ relative to $C_{eq}$ has a large impact on full circuit design and frequency response
as described in \citet[Sect.~4.3]{piat:tel-00004038}.
The stray capacitance occurs from the length of cable between the detector and the JFet Box
of the electronic readout chain of \HFI{} and it applies a low pass filter on the bias current $I$. 
It is a design goal to mimic DC bias~\citep{2010A&A...520A...9L} with a square shape for $I$. 
It allows to have a joule effect as stable as possible and then reduce variations of $R$ which maximize the linearity of the response.
This is why the bias voltage $V_{\rm bias}$ is composed of a triangle plus a {\it compensation} square wave.
The square wave allow to compensate for $C_s$ by stabilizing quickly the current through the bolometer.
The major drawback from the square wave addition is a frequency shape in {$1/f$}, 
while the triangle wave alone has a frequency shape in $1/f^2$.
As an effect, the number of harmonics needed to describe the signal will be significantly increased,
as for the computational complexity.

\subsection{Thermal model}
\label{thermal_model}

\begin{figure}[t!]
    \centering
    \includegraphics[width=0.35\textwidth]{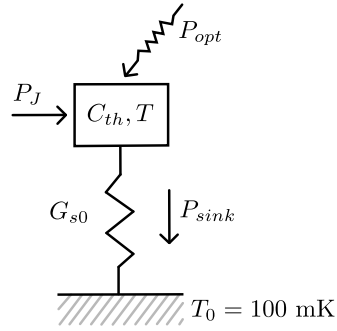}
    \caption{Standard bolometer thermal model. 
    	The bolometer itself is represented by the central box with heat capacity $C_{th}$ and temperature $T$.
	$P_{opt}$ is the total optical radiative power falling on the detector. 
	$P_J$ is the Joule effect power produced by bias current.
	$P_{sink}$ is the power flowing out of the bolometer via the thermal link with static thermal conductance $G_{s0}$.
	The heat sink at temperature $T_0$ is supposed to have an infinite thermal capacity.
	}
    \label{fig_thermal_model}
\end{figure}

The thermal equilibrium of the bolometer is described by the standard theoretical bolometer model. We will
use the conventions proposed by~\citet{Piat2006588}.
The thermal equilibrium, as represented in Fig. \ref{fig_thermal_model}, can be written
\begin{equation}
	C_{th} \frac{dT}{dt} = P_J +\Popt - P_{sink},
	\label{thermal_equi}
\end{equation}
where $C_{th}$ is the bolometer heat capacity, $T$ is the bolometer temperature, $P_J=\frac{V^2}{R}$ is the power dissipated through the bolometer by Joule effect, $P_{sink}$ is the power dissipated through the heat sink, and $\Popt$ is the optical radiative power falling on the bolometer.

The bolometer heat capacity is given as a function of temperature by ~\citep{Mather:84} 
\begin{equation}
	C_{th}(T) = C_0 T^\gamma \, ,
	\label{Cth_of_T}
\end{equation}
where $\gamma$ and $C_0$ are experimental measurements  obtained during calibration campaigns
at Jet Propulsion Laboratory~\citep{2008ApOpt..47.5996H}.
Numerical values are referenced in Table~\ref{table_therm_params}.

In the absence of electric field, the variation of bolometer impedance can be expressed as a function of its temperature 
from the simple bolometer model
\begin{equation}
  R(T) = R_G e^{ \sqrt{ \frac{T_G}{T} } },
  \label{R_of_T}
\end{equation}
where $R_G$ and $T_G$  are experimental measurements.

The power dissipated through the heat sink is a function of bolometer temperature,
it can be written as~\citep{Mather:82} 
\begin{equation}
	P_{sink}(T) = \frac{G_{s0}}{T_{ref}^\beta(\beta+1)}(T^{\beta+1}-T_0^{\beta+1}) \, ,
	\label{Psink_of_T}
\end{equation}
with $\beta$, $G_{s0}$ the static thermal conductance, being experimental measurements.
$T_{ref}$ is a reference temperature and the heat sink temperature is fixed to $T_0=100mK$.


\section{Linearizing differential equations}
\label{sect_linear_versions}

In this section, we describe how electrical and thermal equilibriums can be rewritten to obtain
a linear system of two equations of  V and R only.
We make a clear distinction between steady state and optical excitation signals because we need to solve
for the two cases separately.
For this purpose Taylor expansions are used to reach a target goal of $10^{-3}$ relative precision.
This is a reasonable objective, because as we will see in Sect.~\ref{sect_response_linearity}, the simulations show
the response to an optical excitation is linear at $10^{-5}$ level.

These linearized forms will be used in Sect.~\ref{section_formal} where 
we describe a method to solve the problem in frequency domain.

\subsection{Signals decomposition}

Let start by making a clear distinction between signals in steady state, and the ones generated by optical excitation.
We make the hypothesis that the incoming optical radiative power is the sum of
a constant bacground $\Pavg$ and a small monofrequency optical excitation term $\PR$.

The bolometer response to $\PR$ is linear to a good approximation, so we make the hypothesis that we can write its voltage as
\begin{equation}
  V(t) = \Vavg + \MV(t) + \EV(t) +  \order(\EV^2) \, ,
\end{equation}
using the notations
\begin{itemize}
	\item  $\Vavg$ for the average value of $V$ (in this case $\Vavg=0$)
	\item $\MV$ for the steady state modulation harmonics 
	\item $\EV$ for the harmonics appearing in response to $\PR$
\end{itemize}
The second order terms in $\order(\EV^2)$ can be neglected, see Sect.~\ref{sect_simu} for quantitative values.

The same hypothesis and notations will be applied to $T(t)$, $R(t)$, $C_{th}(t)$, $P_{sink}(t)$ and $P_J(t)$.
These notations allow to distinguish easily the order of the terms in the equations as we
do have in practice  the generic relation $ \Tavg \gg \MT \gg  \ET$. Numerical values will be detaild in Sect.~\ref{sect_simu}.
Consequently, from now, all terms containing a product of two optical excitation component like $\EV \ER$ 
will be considered as negligible, and nonlinear.

\subsection{Electrical equilibrium}
\label{sect_elec_equi}

The electrical equilibrium is already a function of $V$ and $R$.
Using previously defined notations, we separate the steady state
terms from the optical excitation terms. 
Then Eq.~(\ref{electrical_equilibrium}) reads
\begin{equation}
	\MV + \Rsteady (C_{eq}+C_s) \dt{\MV} 
		= 	\Rsteady  C_{eq} \dVbias \, ,
	\label{linear_elec_equi_steady}
\end{equation}
and
\begin{equation}
	 \EV + ( C_{eq} + C_s ) \left( \Rsteady  \dt{\EV} + \ER \dt{\MV} \right)   	
	 		=	\ER C_{eq} \dVbias +\order(\ER \EV) \, .
	 \label{linear_elec_equi}
\end{equation}

\providecommand{\e}{\ensuremath{\textsc{e}}}

\subsection{Thermal equilibrium}
\label{sect_thermal_equi}

The thermal equilibrium Eq.~(\ref{thermal_equi}) is a function of $T$, $V$ and $R$.
In this section, we rewrite its terms as linearized versions of $V$ and $R$ only. 
To do so we first express $T$ as a function of $R$.

\subsubsection{Expressions of $T$ from $R$}

\newcommand{\partialat}[3]		{\ensuremath  \left.  \frac{ \partial #1 }{  \partial #2} \right|_{ #3 }   }

From Eq.~(\ref{R_of_T}) the value of $T$ is
\begin{equation}
	T(R)		=	\frac{ T_g } { \left( \ln R - \ln R_g \right)^2 }
	\label{T_of_R}
\end{equation}

As motivated by simulation analysis (see Sect.~\ref{sect_simu}), the time derivative of optical excitation component has $15\%$ error when
using 1st order Taylor expansion, so we will provide coefficients obtained from Eq.~(\ref{R_of_T}) for the two first orders which reduces error to $0.2\%$. These coefficients reads
\begin{eqnarray}
	\partialat{T}{R}{\Tavg}	&	= a_{1T} =		&  - \frac{ 2 T_g } { \Ravg \left( \ln \Ravg - ln R_g \right)^3 } \, , \cr
	\frac{1}{2!} \partialat{^2 T }{R^2}{\Tavg} & = a_{2T} = &  \frac{ T_g \left( \ln \Ravg -ln R_g + 3 \right) } { R^2 \left( \ln \Ravg - \ln R_g \right)^4 } \, ,
	\label{T_coefs}
\end{eqnarray}
so we can write
\begin{equation}
	T(\Ravg + \delta R) = \Tavg + a_{1T}  \delta R + a_{2T} \delta R^2 + \order(\delta R^3)
\end{equation}

For steady state plus optical excitation component $\delta R = \MR + \ER$, then we can write 
the Taylor expansion of temperature variations as
\begin{align}
	T(\Ravg+\delta R)	= 	\Tavg  + \underbrace{ a_{1T} \MR + a_{2T} \MR^2 }_{\MT} 
						+  \underbrace{ a_{1T} \ER 	+ 2  a_{2T}  \MR  \ER }_{\ET} 	 + \order(\delta R^3)  \,  ,
	\label{T_taylor2}
\end{align}
where the $a_{2T} \ER^2$ term has been neglected in the expression.

And finally the time derivatives of \MT and \ET, which we will need for linear expression of Eq.~(\ref{thermal_equi}), 
can be written from the derivation of Eq.~(\ref{T_taylor2}) over time as 
\begin{eqnarray}
	\dt{\MT}	&	=	& a_{1T} \dt{\MR} + 2 a_{2T} \dt{\MR} \, \MR +\order(\MR^3) \, , \cr
	\dt{\ET}	&	=	& a_{1T} \dt{\ER} + 2 a_{2T}  \left(  \dt{\MR} \ER + \MR \, \dt{\ER} \right) + \order(\ER^3) \, .
	\label{dT_taylor2}
\end{eqnarray}

\subsubsection{Heat quantity variations}

\newcommand{\Pj}	{{\langle P_J \rangle}}
\newcommand{\Cth}[1][1]	{\ensuremath{\langle C \ifthenelse{1=#1}{_{th}}{_{th#1}} \rangle}}
\newcommand{\Psink}[1][1]	{\ensuremath{\langle P \ifthenelse{1=#1}{_{sink}}{_{sink#1}} \rangle}}
\newcommand{ \MCth }  { \ensuremath{\overline{ C}_{th} }}
\newcommand{ \ECth }  { \ensuremath{\widetilde{ C}_{th} }}

We now build the linear version of heat capacity $C_{th}$ defined in Eq.~(\ref{Cth_of_T}) as a function of R.
Its first order Taylor expansion, as a function of $T$,  is
\begin{equation}
	C_{th}(\Tavg + \delta T) =  \Cth +  \underbrace{\partialat{C_{th}}{T}{\Tavg}}_{a_{1C}} \delta T +\order(\delta T^2) \, , 
\end{equation}
with $\Cth = C_0 \Tavg^\gamma$ and $a_{1C}=\gamma C_0 \Tavg^{\gamma-1}$.

To build the expression as a function of $R$ at first order, we can replace $\delta T$ 
by  $a_{1T}\delta R$ using Eq.~(\ref{T_coefs}). Then the 1st order Taylor expansion becomes
\begin{equation}
	C_{th}(\Tavg + \delta T) =  \Cth +  a_{1C} \delta R +\order(\delta R^2) \, , 
\end{equation}
with $a_{1C} = \left( \gamma  C_0  \Tavg  ^{\gamma-1} \right)  a_{1T}$.
Using $\delta R = \MR + \ER$ we can express $C_{th}$ for steady state 
plus an optical excitation as
\begin{equation}
	C_{th}(\Tavg + \delta T) = \Cth + \underbrace{a_{1C} \MR}_{\MCth} 
							  + \underbrace{a_{1C} \ER}_{\ECth} 
							  + \order(\ER^2)  \, .
\end{equation}

Now the linear version of $C_{th} \dt{T}$ term of Eq.~(\ref{thermal_equi})
reads
\begin{equation}
	C_{th} \dt{T} = \left(  \MCth + \ECth \right) \left(  \dt{\MT} + \dt{\ET}  \right) \, .
\end{equation}
We can separate the steady state only term from the product. Using Eq.~(\ref{dT_taylor2}) the result
as a function of $R$ reads
\begin{eqnarray}
	\overline{ \left( C_{th}  \dt{T} \right) }	&	=	&  \MCth \, \dt{\MT} + \order(\MR^5) \cr
		&	=	&
		 \left( \Cth + a_{1C} \MR \right) 
		\left( a_{1T} \dt{\MR} + 2a_{2T}\dt{\MR} \, \MR \right) + \order(\MR^5) \cr
		&	=	& \Cth a_{1T} \dt{\MR} +\order(\MR^2) \, .
	\label{CthdT_steady}
\end{eqnarray}
where the $\MR^2$ terms have been neglected.
The term for an optical excitation is respectively
\begin{eqnarray}
	\widetilde{ \left(  C_{th} \dt{T} \right) }	&=&\MCth \dt{\ET} + \ECth \dt{\MT} + \order(\ECth {\ET})
	\cr
		& = &\left( \Cth + a_{1C} \MR \right) 
		\left( a_{1T} \dt{\ER} + 2 a_{2T} \left( \dt{\MR} \ER + \MR \dt{\ER} \right) \right) 
		+ a_{1C} \ER
		\left(   a_{1T} \dt{\MR} + 2 a_{2T} \dt{\MR} \, \MR  \right) + \order(\ER^2) \, .
	\label{CthdT_excited}
\end{eqnarray}

\subsubsection{Heat sink dissipated power}
\label{sect_heat_sink}

To build the linear expression of the power dissipated trough the heat sink
as a function of $R$, we will consider the first order Taylor expansion
of Eq.~(\ref{Psink_of_T}) given by
\begin{eqnarray}
	P_{sink}(\Tavg+\delta T)	=	\Psink + \underbrace{\partialat{P_{sink}}{T}{\Tavg}}_{G_d} \delta T 
						+ \mathcal O ( \delta T^2 ) \, ,
\end{eqnarray}
with 
\begin{equation}
	G_d = G_{s0} \left( \frac{T}{T_{ref}} \right)^\beta,
\end{equation}
where $G_d$ is the dynamic thermal conductance.

This expression can be written as a function of $R$, by replacing 
$\delta T$ using  Eq.~(\ref{T_taylor2}) as
\begin{align}
  P_{sink}(\Tavg + \delta T) 	=		
  		\Psink +  \underbrace{ G_d a_{1T} \MR }_{\overline{P}_{sink}}
		+ \underbrace{ G_d \left( a_{1T} \ER + 2 a_{2T} \MR \ER \right)}_{\widetilde{P}_{sink}} + \order(\ER^2) \, . 
\label{linear_Psink}
\end{align}

\subsubsection{Joule effect}

By definition

\begin{equation}
  P_J  =  \frac{(\MV + \EV)^2}{\Ravg + \MR+\ER} \, .
\end{equation}

The denominator is of the form $\frac{1}{(x + \epsilon)}$, with $x=\Ravg$ and $\epsilon=\MR+\ER$.
 Its first order Taylor expansion is 
$\frac{1}{x}-\frac{\epsilon}{ x^2}$, which leads to
\begin{equation}
	P_J	=		\underbrace{ \frac{\MV^2}{\Ravg} - \frac{\MV^2 \MR}{\Ravg^2} 
					    }_{\overline{P}_J}
				+ \underbrace{ 2\frac{\MV \EV}{\Ravg} - \frac{ \MV^2 \ER}{\Ravg^2} - \frac{2\MV\,\MR\EV}{\Ravg^2} 
					    }_{\widetilde{P}_J} 
				+ \order(\ER^2) \, . 
    \label{linear_Pj}
\end{equation}


\section{Solving in frequency domain }
\label{section_formal}

In the following section, we build the optical transfer function describing the
detector voltage response to an optical excitation at angular frequency $\omega$.
To build the solution in frequency domain, a matrix formalism is developed
to work on signals projected in custom Fourier domain basis.
The solving is done separately for steady state only first, and then for optical excitation case.

\newcommand{ \vmr }	 { \ensuremath{{\bf  \overline{r}} }}
\newcommand{ \vmt }	 { \ensuremath{{\bf  \overline{t}} }}
\newcommand{ \vmv }	 { \ensuremath{{\bf  \overline{v}} }}
\newcommand{ \vvbias }	 { \ensuremath{{\bf  V_{\rm bias}} } }
\newcommand{ \poptp }	 { \ensuremath{{\bf  \widetilde{p}^+} }}
\newcommand{ \poptm }	 { \ensuremath{{\bf  \widetilde{p}^-} }}
\newcommand{ \voptp }	 { \ensuremath{{\bf  \widetilde{v}^+} }}
\newcommand{ \voptm }	 { \ensuremath{{\bf  \widetilde{v}^-} }}
\newcommand{ \roptp }	 { \ensuremath{{\bf  \widetilde{r}^+} }}
\newcommand{ \roptm }	 { \ensuremath{{\bf  \widetilde{r}^-} }}
\newcommand{ \toptp }	 { \ensuremath{{\bf  \widetilde{t}^+} }}
\newcommand{ \toptm }	 { \ensuremath{{\bf  \widetilde{t}^-} }}
\newcommand{ \MD }	 { \ensuremath{{\bf  D} }}
\newcommand{ \MX }	 { \ensuremath{{\bf  X} }}
\newcommand{ \MY }	 { \ensuremath{{\bf  Y} }}
\newcommand{ \MDp }	 { \ensuremath{{\bf  D^+}} }
\newcommand{ \MDm }	 { \ensuremath{{\bf  D^-}} }
\newcommand{ \MId }	 { \ensuremath{{\bf  I_d}} }
\newcommand{ \TFp }	 { \ensuremath{ {\bm{  \mathcal{F}} }^+ } }
\newcommand{ \TFpc }	 { \ensuremath{ {\bm{  \mathcal{F}} }^{+*} } }
\newcommand{ \TFm }	 { \ensuremath{ {\bm{  \mathcal{F}} }^- } }
\newcommand{ \TFmc }	 { \ensuremath{ {\bm{  \mathcal{F}} }^{-*} } }
\newcommand{ \TF  }	 { \ensuremath{ {\bm{  \mathcal{F}} } } }
\newcommand{ \TFS }	 { \ensuremath{ \mathcal{F}_\Sigma}}
\newcommand{ \sphase }	 { \ensuremath{ \Delta t}}

\newcommand{ \dcomb }	{\ensuremath \diracComb}
\newcommand{ \pcomb }	{\ensuremath \diracComb^+}
\newcommand{ \mcomb }	{\ensuremath \diracComb^-}
\newcommand{ \Ccirc }	{ \ensuremath{{\bm{  \mathcal{C}} }} }

\newcommand{\vect}[1] 		{{\ensuremath{\bf #1}}}
\newcommand{\vsteady}[1] 	{{\ensuremath{\bf \overline{#1}}}}
\newcommand{\vexcitp}[2]	{ {\ensuremath{\bf \widetilde{#1}_{#2}^+}} }
\newcommand{\vexcitn}[2]	{ {\ensuremath{\bf \widetilde{#1}_{#2}^-}} }

\subsection{Matrix formalism}
\label{sect_matrix_formalism}

Now that we have linear version of differential equations we show how we can use
linear algebra to solve the problem.

For the discrete description of signals in frequency domain, the number of considered modulation 
harmonics will be fixed to $n$. 
The vectors in frequency domain for real signals need to be of size $2n+1$ including complex 
conjugates for negative frequencies. 
Best values for $n$ will be discussed in Sect.~\ref{section_nharms}.
Steady state signals are described first, signals resulting from optical excitation must be handled differently.
After the custom frequency vector basis are defined, the time derivative of signals can be expressed in a simple form.

\subsubsection{Steady state frequency vectors}

Let consider a steady state signals, $\overline{S}(t)$.
It is composed only of modulations harmonics and we can write using complex notation
\begin{equation}
	\overline{S}(t) = \sum_{k=-n}^n \vsteady{s}_k e^{j k \Wm t}\, ,
\end{equation}
where $j$ is the imaginary unit, and $\Wm=2\pi f_{mod}$ is the modulation angular frequency.
The steady state vector resulting from the projection of $S(t)$ in Fourier domain is noted $\vsteady{s}$. 
As $\overline{S}(t)$ is a real valued signal, we do have the relation $\vsteady{s}_{-k} = \vsteady{s}_{k}^*$, 
where $z^*$ is the conjugate of $z$. 
This representation is homogenous with the outputs of the fast Fourier transform algorithm.
We note $\diracComb$ the discrete frequency support of $\vsteady{s}$ defining the
Fourier basis of size $2n+1$ for steady state signals description.

The key point to solve electrical and thermal equilibrium is to represent the product
of time domain signals.
Hopefully the convolution theorem states that a product in time domain is equivalent
to circular convolution product in frequency domain. 
Then let consider the product of two steady state signals, $\overline{S}_1(t)$ and $\overline{S}_2(t)$,
with corresponding steady state vectors $\vsteady{ s_1}$ and $\vsteady{ s_2}$.
Their discrete circular convolution product can be written in matrix notation using the circulant matrix of left vector as
\begin{equation}
	\vsteady{s_1} \otimes \vsteady{s_2} = {\bf \Ccirc(s_1)}  \vsteady{s_2} = {\bf \Ccirc(\vsteady{s_2})} \vsteady{s_1} \, ,
\end{equation}
where $\Ccirc(\vsteady{s_1})  \vsteady{{s_2}}$ is the matrix product of the circulant matrix $ {\bf \Ccirc(s_1)}$ and vector $\vsteady{s_2}$.
The result of the product is also in $\dcomb$.

\subsubsection{Optical excitation vectors}
\label{sect_opt_excit_vect}

\begin{figure}[t!]
  \centering
   \includegraphics [width=0.8\textwidth] {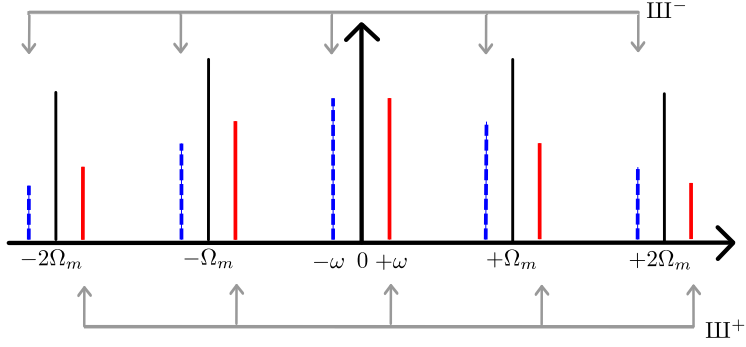} 
    \caption{Schematic for decomposition of a signal $\widetilde{S}_2(t)$ resulting from an optical excitation
    		at frequency $\omega$, with $n=2$.
    		The red solid lines represent the harmonics on the frequency support $\pcomb$.
		While the blue dashed lines represent the harmonics on the frequency support $\mcomb$.
		The size difference around each modulation frequency emphasize that considered modulated harmonics
		result from the system response, and  not from the modulation of a single frequency signal.
		}
    \label{modulated_harmonics}

\end{figure} 

Let consider now a product $\overline{S}_1(t) \widetilde{S}_2(t)$.
The signal $\widetilde{S}_2(t)$ results from the system response to an optical excitation at angular frequency $\omega$.
$\widetilde{S}_2(t)$ is not stricto sensu a modulated signal, because the electro-thermal equilibrium does not build in the form
$\overline{S(t)}*\cos(\omega t + \varphi)$. However its harmonics appears at the same frequencies, as shown in 
Fig.~\ref{modulated_harmonics}.
$\widetilde{S}_2(t)$ can be written
\begin{equation}
	\widetilde{S}_2(t) = \sum_{k=-n}^{n} \left(  \left. \vexcitp{s}{2} \right._k e^{j (k \Wm + \omega) t } + \left. \vexcitn{s}{2}\right._k e^{j (k\Wm - \omega) t } \right)
	\, .
	\label{S2_partition}
\end{equation}
The frequency step between $\widetilde{S}_2(t)$ harmonics is not homogenous. 
Then the convolution product in frequency cannot be written with a single vector for $\widetilde{S}_2(t)$ as for steady state only case.
A classic DFT based method could be used to compute the convolution product. 
However for an exact computation it would require using a frequency step which is a multiple of $\Omega_m$ and $\omega$.
This would render the computing extremely CPU intensive, or infeasible.
Because the solving step would involve inversion
of huge matrix which is an $\order(n^3)$ operation. So we have to use another solution and split the problem in two.

From the coeficients defined in Eq.~\ref{S2_partition}, we can write $\widetilde{S}_2(t) = \widetilde{S}^+_2(t) + \widetilde{S}^-_2(t)$.
The product $\overline{S}_1(t) \widetilde{S}_2(t)$ can be written in frequency domain as
\begin{equation}
	\overline{S}_1(\nu) \otimes \widetilde{S}_2(\nu) = 
		\overline{S}_1(\nu) \otimes \widetilde{S}_2^+(\nu) 
		+ \overline{S}_1(\nu) \otimes \widetilde{S}_2^-(\nu) \, .
\end{equation}
All terms in the right part of the equation have a frequency step of $\Omega_m$.
The frequency support for $\widetilde{S}_2^+$ and  $\widetilde{S}_2^-$ are noted $\pcomb = \llbracket k\Omega_m + \omega \rrbracket$ 
and $\mcomb=\llbracket k\Omega_m - \omega \rrbracket$
with $k \in \llbracket -n \cdots n\rrbracket$.

With this partitioning in frequency, the convolution product can be computed in the same way as for steady state,
but separately for $\vexcitp{s}{2}$ and $\vexcitn{s}{2}$.
Considering that $\vsteady{s_1} \otimes \vexcitp{s}{2}$ builds only terms at frequencies 
$p\Omega_m +(q\Omega_m + \omega) = (p+q)\Omega_m + \omega$, with $p  \in \llbracket-n \cdots n \rrbracket$,
and $q \in  \llbracket-n \cdots n \rrbracket$.
Then it comes
\begin{equation}
	\vsteady{s_1} \otimes \vexcitp{s}{2} \in \pcomb \, .
\end{equation}
Respectively 
\begin{equation}
	\vsteady{s_1} \otimes \vexcitn{s}{2} \in \mcomb \, .
\end{equation}
As a consequence, the terms in $\pcomb$ and $\mcomb$ have to be manipulated separately, 
when working in frequency domain.

As $\widetilde{S}(t)$ is a real valued signal, we do have by construction $\left.\vexcitp{s}{2}\right._{k} = \left.\vexcitn{s}{2}\right.^*_{-k}$ . 
 From this consideration the computing of solutions is needed only for $\vexcitp{s}{2}$ vectors, as the $\vexcitn{s}{2}$ version
 can be built from it.

\subsubsection{Time derivative of vectors}

The time domain derivative of a signal can be written from its DFT vector, considering that 
$\frac{d}{dt} e^{jk \Wm t}=jk\Wm e^{jk\Wm t}$. 
Then the time derivative operation can be represented using a complex square diagonal matrix of size $2n+1$,
and can be written
\begin{equation}
	\frac{d}{dt} {\bf \overline{s}}  = {\bf D} {\bf \overline{s}} \, ,
\end{equation}
with
\begin{equation}
	{\bf D} = 
		\begin{pmatrix}
			-j n \Wm & 0 & 0 \\
			0 & \ddots & 0  \\
			0 & 0    & j n \Wm \\
		\end{pmatrix} .
\end{equation}

Time domain derivatives for optical excitation vectors $\vexcitp{s}{}$ and $\vexcitn{s}{}$ respectively, 
are defined with matrix ${\bf D^+}$ and ${\bf D^-}$ respectively. Their diagonal elements are $j (k \Wm + \omega)$ and  $j (k \Wm - \omega)$ respectively, for $k \in \llbracket-n \cdots n \rrbracket$.

\subsubsection{indexing of vectors}

The $2n+1$ vector elements are referenced by their harmonic index
varying monotonically from $-n$ to $n$. 
The harmonic index $0$ will be considered to be at center of vector.
In practice many DFT implementations store negative elements at the end of the matrix.
This is completely equivalent for the expressions presented in this article.

\subsection{Steady state solving}
\label{sect_steady_state}

The steady state is characterized by $\MV$ and $\MR$, their values must be computed first
as they are needed for solving the optical excitation expression. 
For steady state, the electrical equilibrium Eq.~(\ref{linear_elec_equi_steady}) reads in matrix form
\begin{equation}
	\vmv =  \left( \MId + (C_s+C_{eq}) \Ccirc(\Ravg+\vmr) \MD  \right)^{-1} C_{eq} \Ccirc(\Ravg + \vmr)  \MD \vvbias \, .
	\label{eq_vsteady}
\end{equation}
where $\MId$ is the identity matrix.

Impedance variations $\MR$ are of order $10^{-3}$ of $\Ravg$ (see Sect.~\ref{sect_simu}). 
So the electrical equilibrium behave at first order as a static impedance circuit and
the vector $\vmv$ can be computed directly from Eq.~\ref{eq_vsteady} by setting $\vmr$ to zero.
The value of $\Ravg$ is supposed to be known. 
This is for real life detectors a parameter generally obtained by direct measurement.

The value of the vector $\vmr$ is obtained from thermal equilibrium Eq.~(\ref{thermal_equi}).
With steady state notations thermal equilibrium reads
\begin{equation}
	\overline{ \left( C_{th}  \dt{T} \right) } = \Pj + \overline{P}_J - \Psink - \overline{P}_{sink} \,.
\end{equation}
Using expressions defined in Eq.  (\ref{CthdT_steady}),   (\ref{linear_Psink}) and (\ref{linear_Pj})  
the matrix form is
\begin{equation}
	\vmr		=	\left(  
					\Cth a_{1T} \MD + \frac{ \Ccirc(\Ccirc(\vmv)\vmv) }{\Ravg^2} 
					+ G_d{a_{1T}} \MId
				\right)^{-1} 
				\left(
					\frac{ \Ccirc(\vmv) \vmv}{\Ravg} - \Psink
					+ \Pavg 
				\right) \, ,
	\label{eq_r_steady}
\end{equation} 
where the value of $P_{sink}(\Tavg)$ is obtained using Eq.~(\ref{Psink_of_T}) and (\ref{T_of_R}).
Let's stress here that adding a scalar value like $\Pavg$ to a steady state vector is equivalent to adding
the value to the harmonic index 0 of the vector.

Once the vector $\vmr$ is computed, it can be used to update the value of $\vmv$
by using Eq.~(\ref{eq_vsteady}).
This will improve the precision of $\vmv$ by an order.

A second iteration for $\vmr$ is not usefull because the value used for $\MCth \, \dt{\MT}$ 
in the final version of Eq.~(\ref{CthdT_steady}) cancels second order terms in $\MR^2$. So it would not bring better precision.

\subsection{Optical excitation solving}

Electrical equilibrium equation \ref{linear_elec_equi} reads in matrix form

\begin{align}
	\left\{
	\begin{array}{lcl}
	\roptp 	&	=	&	{\bf E^+} \voptp \\
	\roptm 	&	=	&	{\bf E^-} \voptm \, , \\
	\end{array}
	\right. 
	\label{linear_elec_equi}
\end{align}
with 
\begin{align}
	\left\{
	\begin{array}{lcl}
	{\bf E^+} 	&	=	&	{\bf E1} \left( \MId + (C_{eq}+C_s) \Ravg \MDp \right)  \\
	{\bf E^-} 	&	=	&	{\bf E1} \left( \MId + (C_{eq}+C_s) \Ravg \MDm \right)  \\
	{\bf E1}	&	=	&	\left\{ C_{eq} \Ccirc(\MD {\bf V_{\rm bias}}) - (C_{eq} + C_s) \Ccirc(\MD {\bf \overline{v}} ) ) \right\}^{-1} \, . \\
	\end{array}
	\right. 
\end{align}

The thermal equilibrium Eq.~(\ref{thermal_equi}) reads using optical excitation notations
\begin{equation}
	\widetilde{ \left(  C_{th} \dt{T} \right) }   = \widetilde{P}_J -  \widetilde{P}_{sink} +  \widetilde{P}_{opt} \, .
\end{equation}
Which can be written in matrix form using Eq.~(\ref{CthdT_excited}), 
(\ref{linear_Pj}), (\ref{linear_Psink}) and (\ref{linear_elec_equi}). 
From this expression the analytical transfer function appears in the form of two $2n+1$ square matrices
noted $\TFp$ and $\TFm$ in

\begin{equation}
	\left\{
	\begin{array}{lcl}
	\voptp	&	=	&	 \underbrace{ 
							\left[ 
								\left( {\bf C_1} + {\bf C_2} \MDp - \bf{J_2} + {\bf S} \right) {\bf E^+} - {\bf J_1}
							\right]^{-1} 
						} _{\TFp} 
						\poptp \\
	\voptm	&	=	&	\underbrace{
							\left[
								\left( {\bf C_1} + {\bf C_2} \MDm - \bf{J_2} + {\bf S} \right) {\bf E^-} - {\bf J_1}
							\right]^{-1}
						}_{\TFm} 
						 \poptm , 
	\end{array}
	\right. 
	\label{TF_opt}
\end{equation}
with
\begin{align}
	\left\{
	\begin{array}{lcl}
	{\bf C_1}	&	=	&	\Ccirc \left[ 
							\left\{ (2 \Cth a_{2T}  + a_{1C} a_{1T}) \MId + 2 a_{1C} a_{2T} \Ccirc(\Ravg+\vmr)   \right\} \MD \vmr 
							+ 2 a_{1C} a_{2T} \Ccirc(\MD \vmr ) \vmr
							\right]  \\
	{\bf C_2}	&	=	&	\Ccirc \left[  
							\Cth a_{1T} 
							+ \left\{ (2 \Cth a_{2T}  + a_{1C} a_{1T} ) \MId + 2 a_{1C} a_{2T} \Ccirc( \vmr ) \right\} \vmr 
							\right] \\
	{\bf J_1}	&	=	&	\frac{2}{ \Ravg} \Ccirc( \vmv ) - \frac{2}{\Ravg^2} \Ccirc( \vmv) \vmr  \\
	{\bf J_2}	&	=	&	-\frac{1}{ \Ravg^2} \Ccirc(  \Ccirc(\vmv) \vmv ) \\
	{\bf S}	&	=	&	G_d  \Ccirc \left(   a_{1T} + 2 a_{2T} \vmr  \right)
	\end{array}
	\right. 
	\label{TF_opt_components}
\end{align}
And where $\poptp $ and $\poptm$ describe the, real valued, optical excitation signal 
at angular frequency $\omega$. Considering $\widetilde{P}_{opt}(t) = ze^{j\omega t}+z^*e^{-j\omega t}$, then
$\poptp$ will only have one nonzero coefficient at modulation harmonic index $0$ with value $z$,
respectively $\poptm$ will only have one nonzero coefficient at the same index with value $z^*$. 
${\bf C1}$, ${\bf C2}$ are heat quantity variation matrices.
${\bf J1}$, ${\bf J2}$ are Joule effect matrices.
And ${\bf S}$ is heat sink dissipated power matrix. 
The expression is built in a way to make clearly visible which matrices depends on $\omega$ (with a $+$ or $-$ sign as exponent) 
from the ones that can be computed once and for all.


\section{Results }
\label{sect_results}

In this section we will present the simulation tool used to build the time domain signal response
to optical excitation. Then transfer function results will be validated from the simulation results.
And finally we will discuss the transfer function shape in the data processing low sampling frequency case.

\subsection{Simulation setup }
\label{sect_simu}

The tool used to build realistic simulations conform to \HFI{} readout signal
 is the Simulation for Electronic and Bolometers ({\tt SEB}) which has been developed at IRAP. 
It is an IDL implementation of the electrical and thermal differential equations
of the bolometer with the bias circuit presented in Sect.~\ref{sect_bias}. 
The numerical integration is performed using finite differences with Runge Kuta order 4 method with 10000 points per modulation period.
Simulations have been computed with the setup of the 1st 100GHz channel of \HFI{} ({\tt 00\_100-1a})
with the numerical values presented in Table~\ref{table_elec_params} and Table~\ref{table_therm_params} of
Appendix~\ref{appendix_simu_setup}. 
Simulation outputs have been checked with two other simulations tools~:
{\tt SIMHFI}\footnote{{\tt SIMHFI} is a simulation tool developed with the {\tt LabView} software by R. V. Sudiwala at Cardiff University}
and {\tt SHDet}
\footnote{{\tt SHDet} is a fast simulation tool developed with the {\tt C} programming language by S. R. Hildebrandt at Jet Propulsion Laboratory},
yielding a very good agreement between all approaches.

The simulation setup uses experimental measurements from the \HFI{} first 100GHz channel.
Electrical and thermal parameters are given in Table~\ref{table_elec_params}  and 
Table~\ref{table_therm_params} respectively which are provided in Appendices (Sect.~\ref{appendix_simu_setup}).

Timelines of two seconds are produced, and 1 second of data is discarded to allow the steady state to stabilize
at numerical precision.
The nominal bias voltage is built with a triangle wave of amplitude $vtri$ (not peak to peak) plus a square wave 
of amplitude $vsqu$. Additionally a linear slope of 4.9\% of the modulation period is
added on the square wave at up and down location to mimic real signal raise and fall time.

Two runs of {\tt SEB} have been defined~:
\begin{enumerate}
	\item the reference for steady state with a constant optical background $\Popt = \Pavg$;
	\item the response to an optical excitation with $\Popt = \Pavg + \widetilde{P}_{opt}$,
	 where $\widetilde{P}_{opt}$ is a sin wave of amplitude $9,6593 . 10^{-18}{\rm W}$  (about 3\% of CMB dipole from Doppler effect due to solar system motion) and angular frequency $\omega=\Wm / 18$ which is close to $10{\rm Hz}$;
\end{enumerate}
see Table~\ref{table_numval} in Appendix~\ref{sect_num_values} for amplitudes of $\ER$, $\ET$ and $\EV$ signals
relative to their average and steady state values.

\begin{figure}[t!]
  \centering
   \includegraphics [width=0.7\textwidth] {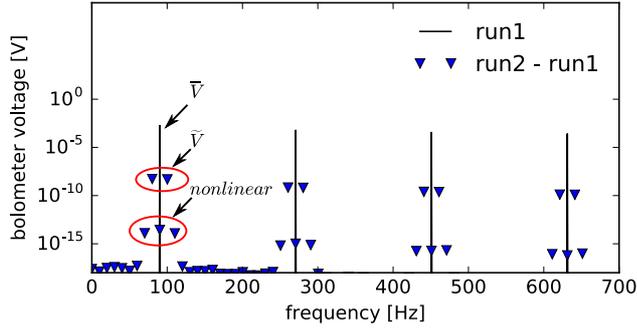} 
    \caption{Bolometer optical excitation linearity.
    		Black lines are bolometer voltage for run1, only the modulation harmonics are visible.
    		Blue triangles are the excitation residuals of bolometer voltage for run2 - run1. 
		Optical excitation signal $\EV$ appear at frequencies $ (2k+1)\Wm \pm \omega$. 
		Nonlinear excitation harmonics $\order(\EV^2)$ are visible 
		about 5 orders bellow $\EV$ at frequencies $(2k+1)\Wm \pm 2\omega$. 
		A modulation nonlinear response  is visible at the same level as for $\order(\EV^2)$ 
		at frequencies $(2k+1)\Wm$.} 
    \label{simu_fft_diffs}

\end{figure}


\subsection{Validation of response linearity}
\label{sect_response_linearity}

First we check the linearity of the system response to an optical excitation. 
The value chosen for $\omega$ in run 1 and 2 allow for the optical period to cover exactly 18 modulation periods,
so the frequencies of interest are not aliased. 
Simulation results are shown in frequency in  Fig. \ref{simu_fft_diffs}. 
The optical response $\EV+\order(\EV^2)$ is produced with the output voltage of ${\rm run}2-{\rm run}1$ 
and is about 5 orders bellow the main modulation harmonics.  
Optical excitation harmonics appear at $(2k+1) \Wm \pm \omega$.
Nonlinear response components $\order(\EV^2)$ appear at $(2k+1) \Wm \pm 2\omega$ and are about 5 orders 
bellow $\EV$.

So the theoretical bolometer model performs very well in the
simulation setup with nonlinear behavior at $10^{-5}$ level of optical excitation response.
This result is in good agreement with inflight results from planet crossings estimated at $10^{-4}$ level, this topic is discussed
in section 3.4 of  \citet{2014A&A...571A...7P}.

\subsection{Transfer function performances}
\label{section_nharms}


Next, we check the convergency of the analytical model as a function of the number of modulation harmonics $n$.
When solving the system of equations in matrix form, there is a competition between 
the number of harmonics increasing precision, and the matrix conditioning increasing the systematic error
due to the finite frequency support and the $1/f$ signal shape.
As a consequence the error as a function of $n$ should reach a plateau then increase again.

\begin{figure}[t!]
    \centering
    \includegraphics[width=0.7\textwidth]{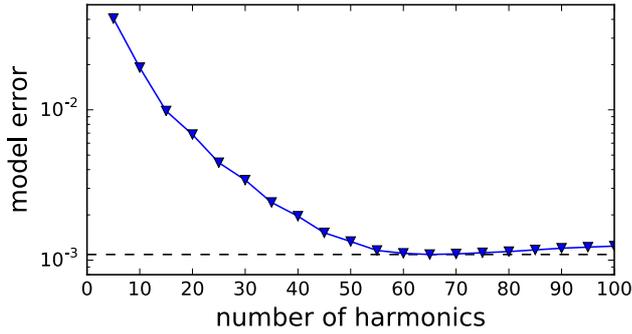}
    \caption{Convergency of the analytical transfer function time domain residuals as a function of $n$.
    	Blue triangles are the relative error of the analytical transfer function output compared to the ${\rm run}2-{\rm run}1$ reference signal.
	The black dashed line is drawn at the minimal error value which is about $10^{-3}$.
    }
    \label{nharms_vo_err}
\end{figure}

\begin{figure}[t!]
    \centering
    \includegraphics[width=0.7\textwidth]{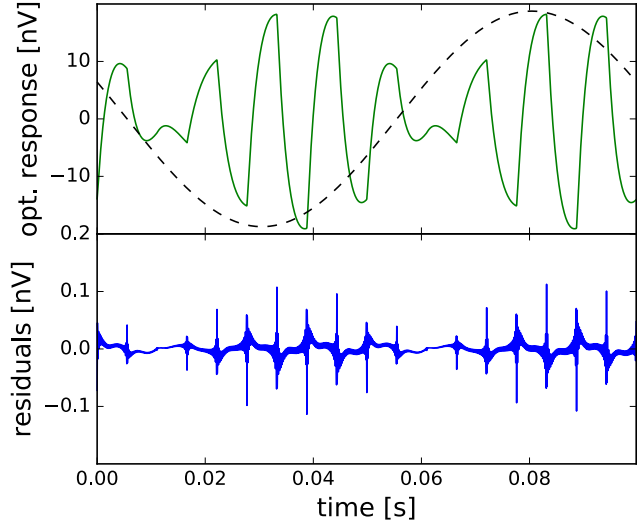}
    \caption{Time domain response to the 10Hz optical excitation.
    		The upper plot green line is the $\EV+\order(\EV^2)$ signal obtained from run2-run1, and referenced as $\EV_{\rm simu}$ in text.
		The dashed black sinusoid is the scaled input optical excitation.
		The bottom plot show the residual error between the simulation and analytical transfer function output
		and referenced as $\EV_{\TF}-\EV_{\rm simu}$ in text.
		}
    \label{time_vo}
\end{figure}

The model error is computed in time domain with the expression $\sigma(\EV_{\TF}-\EV_{\rm simu})/\sigma(\EV_{\rm simu})$,
where $\EV_{\TF}$ is the optical response computed from the analytical model and $\EV_{\rm simu}$ is
computed from simulation data $\rm{run}2-\rm{run}1$ .
The convergency as a function of $n$ is shown in Fig.~\ref{nharms_vo_err}, and the optical response in time domain is shown in Fig.\ref{time_vo} for the optimal value of $n$.
The lowest relative error for $\voptp$ is $10^{-3}$ obtained with $65$ harmonics.
This result is in very good agreement with the target error objective and the measured
error from linear version of equilibrium equations as shown in Table~\ref{table_numval}.
Depending on the target goal, as few as 15 harmonics are necessary to reach a 1\% precision objective.

\subsection{Response for $w \rightarrow 0$ }
\label{sect_raw_gain}

\newcommand{\RG}		{\overline{G}}
\newcommand{\RC}		{\overline{C}}
\newcommand{\vmg}	{\overline{{\bf g}}}

For $\omega=0$, we call {\em steady state gain} the response to  a small change $\delta \Popt$ in the 
constant optical load $\Pavg$ falling on the detector.
We will see in this section how it can be inferred from the analytical model matrix expression,
and in which frequency domain it can be used for slowly varying signal model.

The {\em steady state gain} is an observable of particular importance~:
\begin{itemize}
	\item it provides a calibration source given the knowledge of bolometer thermal and electrical parameters;
	\item it can provide constraints on the bolometer optical response;
	\item it allows us to build a first order model of electronic response shape for slowly varying signal.
\end{itemize}
The later has been used for the ADC nonlinearity correction of \HFI{} data \citep{2015arXiv150201586P}.

\begin{figure}[t!]
    \centering
    \includegraphics[width=0.7\textwidth]{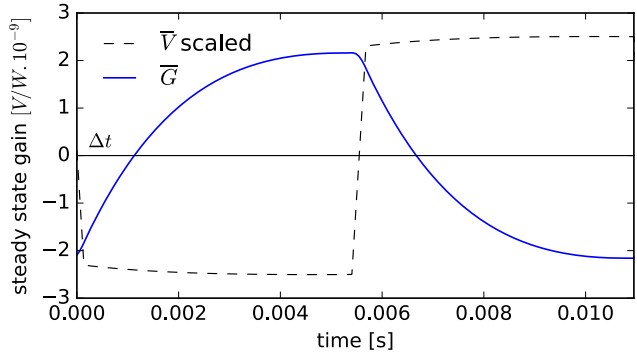}
    \caption{ steady state gain $\RG(t)$ for $\omega=0$ built using run1 configuration and $n=35$.
    		One period is shown. $\sphase$ is the location when $\RG(t)$ changes sign and is
    		a tuning parameter of the integrated version of the transfer function, see Appendix \ref{appendix_TFS}. }
    \label{dynamic_gain}
\end{figure}

Starting from the matrix expression Eq.~(\ref{TF_opt}), $\voptp = \TFp(\omega) \poptp$ and $\voptm = \TFm(\omega) \poptm$.
We notice that only coefficient of $\poptp$ and $\poptm$ at harmonic index zero are non null.
Then only one column of the $2n+1$ square matrices $\TFp$ and $\TFm$ at harmonic index 0 are needed to describe
the optical response. Expanding the time domain expression of $(\voptp, \voptm)$ reads
\begin{equation}
	\left. \EV(t) \right|_\omega = p \sum_{k=-n}^n \TFp_{k,0} e^{j(k\Wm+\omega)t} + p^* \sum_{k=-n}^n \TFp_{k,0} e^{j(k\Wm-\omega)t} \, .
\end{equation}
We have $\TFp(0) = \TFm(0)$, and also $\TFp_{k,0}=\TFmc_{-k,0}$ by construction property of optical excitation vectors as seen
in Sect.~\ref{sect_opt_excit_vect}. Then we can write $\EV(t)$ as a product
\begin{align}
	\left. \EV(t) \right|_{\omega=0}	&	=( p e^{j\omega t} + p^*e^{-j\omega t} )  
									\sum_{-n}^n \left( \TFp_{k,0} e^{jk\Wm t} + \TFpc_{k,0} e^{-jk\Wm t}  \right) & \cr
							&	= \widetilde{P}_{opt}(t) \RG(t) \, , & 
\end{align}
where $\RG (t)$ is the {\em steady state gain} with the same periodicity as modulation,
 and $\vmg_{k} = \TFp_{k,0}(0) + \TFm_{k,0}(0) = 2 \TFp_{k,0}(0) $.  The {\em steady state gain} period is shown for the run1 setup in Fig. \ref{dynamic_gain}.
It appears here that $\RG(t)$ is very different in shape from $\MV(t)$, also the half period signs are opposite because the
impedance variations are negative for a positive change in temperature of the bolometer.

We have seen that for $\omega=0$ we have $\EV(t) = \widetilde{P}_{opt}(t) \RG(t)$.
This is the expression used for slowly varying signal model.
Now we want to characterize its robustness for $\omega \rightarrow 0$ .
The values of $\TFp(\omega)$ and $\TFm(\omega)$ are different when $\omega > 0$ and lead to an expression
who drifts from the product of two time domains signals. 
As $\RG(t)$ is built using the first column of $\TFp(0)$, a simple heuristic is to compare
$\vmg$ with the first columns of $\TFp(\omega)$ and $\TFm(\omega)$.
The Parseval's theorem allowing us to switch the comparison to frequency domain.

\begin{figure}[t!]
    \centering
    \includegraphics[width=0.7\textwidth]{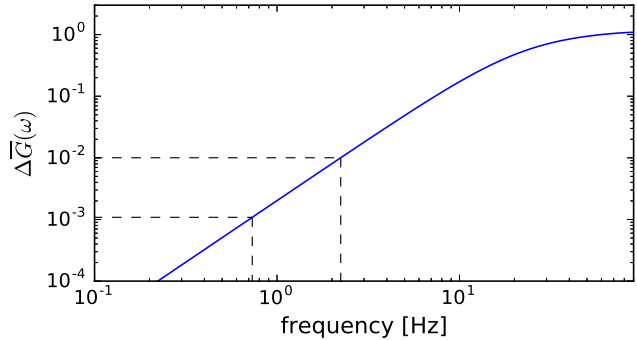}
    \caption{ Slowly varying signal hypothesis error as a function of optical frequency. 
    	The dashed lines indicate locations where the estimated error level is 1\% and 0.1\%.
	The frequency range is bounded on the right at modulation frequency.}
    \label{fig_delta_g}
\end{figure}

We  estimate the slowly varying signal hypothesis error with
\begin{equation}
	\Delta \RG(\omega) = 
		\sqrt{ 
			\frac{  
				\sum\limits_{k=-n}^n   \left|  \vmg_{k} - \TFp_{k,0} (\omega) -\TFm_{k,0} (\omega) \right|^2  
			} {   
				\sum\limits_{k=-n}^n \left| \vmg_{k} \right|^2  
			} 
		}  \, .
\end{equation}
The result is shown in Fig. \ref{fig_delta_g}. 
The slowly varying signal hypothesis  $\EV(t) = \widetilde{P}_{opt}(t) \RG(t)$  exhibits less than 1\% 
estimated error for $\omega < 2 Hz$.

\subsection{Integrated version}

\begin{figure}[t!]
    \centering
    \includegraphics[width=0.7\textwidth]{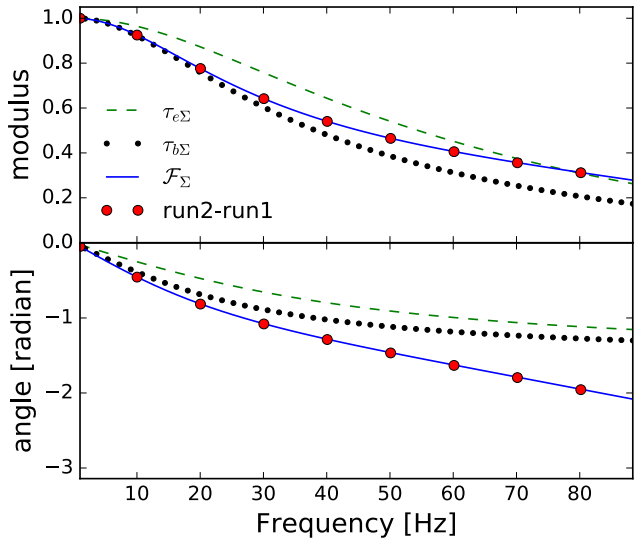}
    \caption{ Integrated version $\TFS$ of the transfer function in the modulation frequency range.
	Upper plot is the normalized modulus at 0 frequency, bottom plot is for the angle in radian.
	Plain blue line is for $\TFS$, red dots are the values computed 
	using (run1,run2) for validation (average relative error of $1,9.10^{-3}$), the dashed green curve is for $\tau_e=4.1ms$ and the black dots are for $\tau_b=6.5ms$ }
    \label{tf_integ}
\end{figure}

We have inspected the performances of the analytical model of the transfer function performances at high frequencies.
Now we check it's behavior in the \HFI{} data processing common use case.
The 80 samples per modulation period are summed over each half period before being send to earth and demodulated.
The integrated version $\TFS(\omega)$, including summation on 40 samples and demodulation,
is extensively described in Appendix $\ref{appendix_TFS}$.
The output is shown in Fig. \ref{tf_integ} for the modulation frequency range. 

For numerical validation, the summation and demodulation operation has been
run on the output of {\tt SEB} for run1 and run2, and is shown as red dots on Fig. \ref{tf_integ}.
With an average difference of $1.9 \times 10^{-3}$ between the simulation and the analytical expression 
of $\TFS$ the outputs is perfectly in the range of the linear approximation used.

For comparison purposes, the integrated version $\tau_{b\Sigma}(\omega)$ and
$\tau_{e\Sigma}(\omega)$ of two low pass filters is also shown.
The generic expression of $\tau_\Sigma(\omega)$ is detailed in appendix \ref{appendix_TFS}.
In the litterature~\citep{Chanin:84,Grannan_Numopt} the bolometer physical time constant is 
under DC bias $\tau_{b}=\frac{\Cth}{G_d}$ and the effective time constant, taking into account the heat flow
from Joule effect, is $\tau_{e} = \frac{\Cth}{G_d - \alpha P_j}$ with  the dimensionless temperature coefficient
$\alpha = \frac{1}{R}\frac{\partial R}{\partial T}$ and in our case, using Eq.~(\ref{T_coefs}), $\alpha= \frac{1}{\Ravg a_{1T}}$.

The AC biased version is significantly different from a low pass filter. 
The $\TFS$ modulus behaves like $\tau_b$ at low frequency and like $\tau_e$ near the modulation frequency,
and the variations of the angle are also more important than for low pass filters.
The observed shape of the analytical model which is a feature specific to AC biasing could explain
why the \HFI{} time transfer function is built with as many as 7 low pass filters~\citep{2015arXiv150201586P}
for some channels.


\section{Conclusions }

We have designed a method to build the analytical expression in frequency domain of the AC biased bolometers response.
Starting from the simple bolometer model used in \HFI{} and the bias circuit of the detector, 
we provided linearized version of the electro thermal equilibrium. 
The target objective is 0.1\% relative precision in presence of a monofrequency optical excitation.
The system of equation is solved using linear algebra using a custom and compact Fourier basis designed
specifically for this purpose.
The method has been applied and tested with the case of the simple bolometer model used in \HFI{}.

The analytical model performances have been characterized using the time domain simulation tool developed at {\em IRAP}.
The reached accuracy is 0.1\% relative to the time domain simulation reference,
when using $n=65$ harmonics.
The solution is built using square matrices of size $2*n+1$ considering a bias signal with $n$ harmonics.
An accuracy of 1\% can be reached using only $n=15$ harmonics.

The proposed analytical model is suitable for deconvolution of real data, as only classical matrix inversion tools are needed.
We also show how the analytical transfer function can be used to build the {\em steady state gain} in time domain.
	This observable provides a first order model of the electronic response, with an accuracy
	of 1\% for optical excitation frequencies less than $2$Hz.
The {\em steady state gain} observable can also be used as a calibration source.

Using the presented matrix formalism, the matrix expression of the transfer
	function allows us to adapt it to different electronic or thermal models.
	A direct extension is to add several thermal components.
	We provide an example with one thermal component.
While the \HFI{} low frequency components of time transfer function still need some improvement when writing this article. 
	The generic form of the proposed model is designed as a tool which can be used to get  better constraints through 
	the measured optical response and the completely new addition of the {\em steady state gain}.
Finally, the analytical transfer function model allow to build a signal model with high resolution in time domain.
	This property is needed to build an improved ADC nonlinearity correction for \HFI{},
	see \citet{2016_ADC_SPIE_ASTO} (SPIE proceeding in preparation).


\newcommand{\Wmp}		{\ensuremath \Omega^{+} }
\newcommand{\Wmm}		{\ensuremath \Omega^{-}  }

\def\appendixpage{\vspace*{8cm} 
\begin{center} 
\Huge\textbf{Appendices} 
\end{center} 
} 

\def\appendixname{Appendix}%

\begin{appendices}

\section{Integrated  transfer function}
\label{appendix_TFS}
 
This appendix describes the integration/downsampling of the electronic signal as done by the onboard Data Processing Unit 
of \HFI{}. The corresponding integrated version of the transfer function $\TFS(\omega)$ will be
built from the matrix formalism described in this paper.
The filtering applied by the readout amplification chain and the rejection filter is
described in the \HFI{} timeresponse paper by~\citet{2014A&A...571A...7P} and will not be considered here.

The data sent to earth at rate $2f_{mod}= 180.3737$~Hz is the sum of $N=40$ samples per modulation half period. 
And the real onboard data acquisition frequency is $2 f_{mod} *N = 7214.948$~Hz before the Data Processing Unit make the summation.
We will start by applying the summation process to a sinusoidal wave described by
\begin{equation}
	s(t) = e^{j\omega t} \, ,
\end{equation}
where $t$ has the same origin as the filtered optical signal.

The sample $i$ acquired at time $t_i=\frac{i}{f_{acq}}$ is
 \begin{align}
 	S_\omega[i] 	&	=	\sum_{k=0}^{N-1}  e^{ j{\omega \left( \frac{i + k/N}{f_{acq} } + \sphase \right) } } \\
				&	=	e^{j \omega \left( \frac{i}{f_{acq}} + \sphase \right) }  \sum_{k=0}^{N-1} e^{j \omega \frac{k}{N f_{acq}}}	\, ,
 \end{align}
 where the time delay term $\sphase$ is a tunable parameter (referenced as $S_{phase}$ by~\citet{2014A&A...571A...7P}) allowing to maximize $\TFS$ gain by adjusting the integration range on the 
 {\em steady state gain} period. As can be seen on Fig. \ref{dynamic_gain} it has a phase advance of 
 about $1/8$ of a modulation period.
 
Considering the sum of the numbers in a geometric progression
\begin{equation}
  \label{geom_progression}
  \sum_{k=0}^{N-1} x^k = \frac{1-x^N}{1-x} \, ,
\end{equation}
then with $x=e^{j \frac{\omega}{Nf_{acq}} }$ the expression of $S_\omega[i]$ can be rewritten
 \begin{align}
 	S_\omega[i] 	&	=	e^{j \omega \left( \frac{i}{f_{acq}} + \sphase\right) } \,  \frac{1-e^{j \frac{\omega}{f_{acq}}}}{1-e^{j \frac{\omega}{Nf_{acq}}}}	\, .
 \end{align}
We  use the Euler relation $\sin(x)= (e^{jx}-e^{-jx})/(2j)$ to simplify the geometric progression sum
 \begin{align}
 	S_\omega[i] 	&	=	e^{j \omega \left( \frac{i}{f_{acq}} + \sphase \right) } \,  
			\frac{
				e^{j \frac{\omega}{2f_{acq}} } \left(e^{-j \frac{\omega}{2f_{acq}} } -  e^{j \frac{\omega}{2f_{acq}} }\right)
			}{
				e^{ \frac{\omega}{2Nf_{acq}} } \left(e^{-j \frac{\omega}{2Nf_{acq}} } -  e^{j \frac{\omega}{2Nf_{acq}} } \right)
			}	\cr
				&	=	e^{j \omega \left( \frac{i}{f_{acq}} + \frac{N-1}{2Nf_{acq}}  + \sphase \right) } \,
			\frac{
				 \sin\left( \frac{\omega}{2f_{acq}} \right)
			}{
				 \sin\left( \frac{\omega}{2Nf_{acq}} \right)
			}	
			\, .
 \end{align}

The term with a sinus denominator is continuous for $\omega \rightarrow 0$ and it's limit is $N$, which can be also
found by using the summation method on a constant signal of value 1 for $\omega=0$. 
The fraction is also continuous for $\omega = kN\Wm , k \in \mathbb{N}$, but these values are out of the
frequency range of interest because they are cut off by the onboard anti aliasing electronic rejection filter.

The sample capture starts at time $t_i$, so the integration transfer function reads
\begin{equation}
	\Sigma(\omega) = e^{j \omega \Delta t_I } 
			\frac{
				 \sin\left( \frac{\omega}{2f_{acq}} \right)
			}{
				 \sin\left( \frac{\omega}{2Nf_{acq}} \right)
			} \, ,
	\label{eq_TFI}
\end{equation}
with $\Delta t_I =\frac{N-1}{2Nf_{acq}}  + \sphase$.

The following step involve the folding of bolometer voltage response built
using $\TFp$ and $\TFm$ matrices.
The modulated linear response appears at odd modulation harmonics.
We will consider only these significant harmonics
at angular frequency  $(2*p+1) \Wm$, and we will use the notation $\Wmp_k = k\Wm+\omega$.

Let consider the complex optical excitation power on the bolometer
\begin{equation}
	\widetilde{P}(t)=e^{j \omega t} \, .
\end{equation}
then, using the matrix formalism developped in Sect.~\ref{sect_matrix_formalism} only one vector is needed 
to represent it with $\poptp = [1, 0 \cdots]$. And the complex output voltage is
\begin{equation}
	\EV(t) = \sum_{\substack{p \ge -\frac{n+1}{2} \\k=2p+1}}^{p \le \frac{n-1}{2}} 
		    \TFp_{k,0} e^{j \Wmp_k t}  \, .
\end{equation}
The acquisition of summed samples at time $t_i$ for odd modulation harmonics folds the signals so
$z e^{j \Wmp_{k} t_i} = z^* e^{j (\Wm-\omega) t_i}$. The folded version of output voltage
at $f_{acq}$ sampling frequency is
\begin{equation}
	\EV(t_i) = \sum_{\substack{p \ge -\frac{n+1}{2} \\k=2p+1}}^{p \le \frac{n-1}{2}}  
			\TFpc_{k,0} \Sigma^*(\Wmp_k) 
		e^{j(\Wm-\omega)t_i} \, .
\end{equation}
The signal is demodulated by applying a $e^{-j\Wm t}$ factor and taking the conjugate of the result
to get a positive frequency, so we have
\begin{equation}
	\EV_d(t_i) =  
	    \sum_{\substack{p \ge -\frac{n+1}{2} \\k=2p+1}}^{p \le \frac{n-1}{2}}  
		\TFp_{k,0} \Sigma(\Wmp_k) 
	e^{j\omega t_i} \, .
	\label{eq_TFSigma}
\end{equation}

The signal described with this expression integrates $N$ samples from time $t_i$ 
and has a time offset of $\Delta t_I$ appearing in Eq.~(\ref{eq_TFI}) compared to input signal.
The time offset can be corrected to obtain the final causal version of the integrated
transfer function
\begin{equation}
  \TFS(\omega) = 
	    \sum_{\substack{p \ge -\frac{n+1}{2} \\k=2p+1}}^{p \le \frac{n-1}{2}}  
		\TFp_{k,0} \Sigma(\Wmp_k)
		e^{-j\omega \Delta t_I} \, .
	\label{eq_TFS}
\end{equation}

The bolometer under DC current bias behaves as an order 1 low pass filter which can be written
\begin{equation}
  T(w) = \frac{1}{1+j\omega \tau} \, , 
\end{equation}
where $\tau$ is the time constant of the filter.
For comparison purposes we can apply the sampling summation process on $T(w)$.
There is no demodulation in this case and we have
$\tau_\Sigma(\omega) = \Sigma(\omega)T(\omega) e^{-j\omega \Delta t_I}$.
As there is no modulated harmonics the time offset $\Delta t_I$ cancels with the one in $\Sigma(\omega)$ 
so we have
\begin{equation}
	\tau_\Sigma(\omega) = T(\omega) 
	\frac{
	    \sin\left( \frac{\omega}{2f_{acq}} \right)
	}{
	    \sin\left( \frac{\omega}{2Nf_{acq}} \right)
	} \, .
\end{equation}


\section{Adding a thermal component}

We have seen how to compute the bolometer response with only one thermal component.
As it has been seen in \HFI{} with the low frequency excess response~\citep{2015arXiv150201586P}, 
the detector behaves as if there are several thermal component  on the thermal path between the bolometer and heat sink,
which alter significantly the filtering of optical signal.
To represent a more realistic system, we will detail an example of adding one thermal component to the expression of $\TFp$ and $\TFm$.

\subsection{Extended thermal model}

\begin{figure}[!t]
  \centering
   \includegraphics [width=0.55\textwidth] {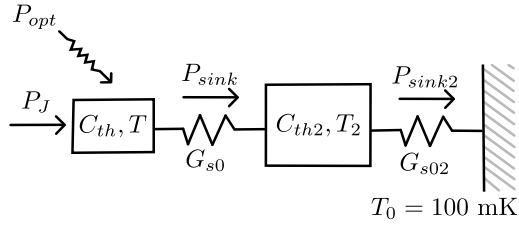} 
    \caption{Extended thermal model with an intermediate heat capacity between the bolometer and the heat sink.
    	The bolometer itself is represented by the left component with thermal capacity $C_{th}$
	and temperature T. $P_J$ is the Joule effect thermal power produced by bias current.
	$P_{opt}$ is the total incoming radiative power falling on the detector.
	$P_{sink}$ is the thermal power flowing out of the bolometer to the second component via
	a thermal link with static thermal conductance $G_{s0}$.
	$C_{th2}$ and $T_2$ are the thermal capacity and temperature of the second thermal component.
	$P_{sink2}$ is the thermal power flowing from the second component via a thermal link
	with static thermal conductance $G_{s02}$. 
	$T_{0}$ is the heat sink temperature with supposed infinite thermal capacity.  }
    \label{fig_thermal_model2}

\end{figure} 

We will use a simple thermal architecture as presented in Fig. \ref{fig_thermal_model2}, by adding
a single component with heat capacity $C_{th2}$ and tempearture $T_2$ between the bolometer and the heat sink.
The new component is connected to the heat sink via a link with thermal conductance $G_{s02}$.

To keep things simple we make the hypothesis that all physical characteristics are well known,
and that for steady state $\Ravg$, $\Tavg$ and $\Tavg[2]$ are also known. In practice, a simple
way to obtain their values is with a fit. If we notice that in steady state $\Psink - \Psink[2]=0$ and that $\Pj+\langle \Popt \rangle-\Psink=0$.
Setting as constraint $T > T_2 > T_0$ we can use the heat flow equilibrium to fit $\Tavg$ anf $\Tavg[2]$.
A commonly used tool as {\tt mpfit}\footnote{
	{\tt mpfit} is a tool for non linear least squares fitting developed by Craig Markwardt and based on the MINPACK-1 software. 
	\url{https://www.physics.wisc.edu/~craigm/idl/cmpfit.html} } 
converge in 5 iterations with a $10^{-5}$ relative difference for the stoppping criteria. 
This is a very quick operation  considering an \order(n) complexity
coming from the $\Pj$ term which needs the computation of voltage harmonics at first order.

Using the same notations as we did in Sect.~\ref{sect_heat_sink}, the heat flow at the output of the 
bolometer can be written at first order 
\begin{align}
	P_{sink} 	= 	\Psink + G_{d12}\MT_1-G_{d21}\MT_2 + \order(\MT^2) + \order(\MT_2^2) \, ,
\end{align}
with	
\begin{align}
\begin{array}{lclcl}
	G_{d12}	&=&	\partialat{P_{sink}}{T}{T=\Tavg}		&=&	\frac{G_{s01}}{T_{ref}^\beta} \Tavg^\beta		\cr
	G_{d21}	&=&	-\partialat{P_{sink}}{T_2}{T_2=\Tavg[2]}	&=&	\frac{G_{s01}}{T_{ref}^\beta} \Tavg[2]^\beta	\, .
\end{array}
\end{align}

The second thermal component on the thermal path can be characterized at first order by 
\begin{align}
\begin{array}{lcl}
C_{th2}		&= 	C_{02} T_2^{\gamma_2}	=	 \Cth[2] + \order(\MT_2)	\cr
P_{sink2}		&= 	\Psink + G_{d2} \MT_2 + \order(\MT_2^2)	\, ,
\end{array}
\end{align}
with
\begin{align}
G_{d2}		&=	\frac{\partial P_{sink2}}{\partial T_2}	=	\frac{ G_{s02} \Tavg[2]^{\beta_2} }{T_{ref}^{\beta_2}}						\, ,
\end{align}
where $\Cth[2]$,$G_{s02}$ and $\beta_2$ are free parameters. We neglect $\Cth[2]$ variations for the following developments.

And the thermal equilibrium for the second component reads at first order
\begin{align}
	C_{th2} \dt{T_2}	&=	P_{sink} - P_{sink2} \cr
	\Cth[2]  \dt{\MT_2}		&=	G_{d12} \MT - (G_{d21} + G_{d2}) \MT_2	+ \order(\MT) \, .
	\label{eq_thermal_equi2}
\end{align}

\subsection{Extended steady state}

The first iteration for $\vmv$ is the same as in Sect.~\ref{sect_steady_state} because we only need to know $\Ravg$.

With the second component, its steady state temperature vector $\vmt_2$ is needed for the thermal equilibrium of
the bolometer.  $\vmt_2$ can be expressed from $\vmr$ using the thermal equilibrium Eq.~(\ref{eq_thermal_equi2}),
then it comes
\begin{align}
	\vmt_2 &= \MX_2 \vmr			\cr
	\MX_2 &= \left[  \Cth[2]	\MD +(G_{d21}+G_{d2}) \MId \right]^{-1} G_{d12} a_{1T} \, ,
	\label{eq_t2}
\end{align}
where $\vmt_2$ is the intermediate steady state vector for $\MT_2$.

And finally the expression of $\vmr$ can be written from the single thermal component version
Eq.~(\ref{eq_r_steady}) by updating the 
$P_{sink}$ term
\begin{align}
	\vmr		=	\left[  
						\Cth a_{1T} \MD + \frac{ \Ccirc(\Ccirc(\vmv).dot(\vmv)) }{\Ravg^2} 
						+ G_{d12} {a_{1T}} \MId -G_{d21} \MX_2
				\right]^{-1} 
				\left[ 
						\frac{ \Ccirc(\vmv) \vmv}{\Ravg} - \Psink 
						+ \Pavg 
				\right] \, ,
	\label{eq_r_steady2}
\end{align} 

Once $\vmr$ is obtained, a second iteration can be done as in Sect.~\ref{sect_steady_state} 
to get $\vmv$ with a better precision.

\subsection{Extended transfer function}

To compute the optical excitation response with a second thermal component, we will make some (optional) simplifications
on the second component. The main hypothesis is that the additional thermal component add a small thermal feedback 
from $\ET_2$ to the heat quantity variations on the bolometer itself. Then we will use the first order expression for heat 
quantity variation, and with the expression of $\ET$ from Eq. \ref{T_taylor2}  the thermal equilibrium will read
\begin{equation}
	\Cth[2] \dt{\ET_2} = G_{d12} \left(  a_{1T} \ER +2 a_{2T} \MR \ER \right) -\left( G_{d21} + G_{d2} \right) \ET_2 + \order(\ER^2) \, .
\end{equation}
From which we can write the vector version of $\ET_2$ as
\begin{align}
\left\{
	\begin{array}{lcl}
		\toptp_2 	&=&	\MY^+_2 \roptp			\cr
		\toptm_2 	&=&	\MY^-_2 \roptm			\, ,
	\end{array}
\right.
\end{align}			
with
\begin{align}
\left\{
	\begin{array}{lcl}			
		\MY^+_2 	&=&	
			\left[ 
				\Cth[2] \MDp 
				+(G_{d21}+G_{d2}) \MId
			\right]^{-1} 
			G_{d12} 
			\left[ 
				a_{1T}\MId 
				+2a_{2T}\Ccirc(\vmr)  
			\right]
		 \cr
		\MY^-_2 	&=&	
			\left[ 
				\Cth[2] \MDm 
				+(G_{d21}+G_{d2}) \MId
			\right]^{-1} 
			G_{d12} 
			\left[ 
				a_{1T}\MId 
				+2a_{2T}\Ccirc(\vmr)  
			\right]	\, .
	\end{array}
\right.
\end{align}

Finally the new transfer function matrices expressions, $\TFp_2$ and $\TFm_2$,  can be written from Eq.~(\ref{TF_opt})
and Eq.~(\ref{TF_opt_components})
by replacing the $P_{sink}$ component ${\bf S}$ with a new version taking into account
the new thermal path
\begin{align}
\left\{
\begin{array}{lcl}
	\TFp_2	&=	&	\left[ 
								\left( {\bf C_1} + {\bf C_2} \MDp - \bf{J_2} + {\bf S_2^+} \right) {\bf E^+} - {\bf J_1}
							\right]^{-1}		\\
	\TFm_2	&=	&	\left[
								\left( {\bf C_1} + {\bf C_2} \MDm - \bf{J_2} + {\bf S_2^-} \right) {\bf E^-} - {\bf J_1}
						\right]^{-1}		\, ,
\end{array}
\right.
\end{align}
with 
\begin{align}
\left\{
\begin{array}{lcl}
	{\bf S_2^+}	&=	&	G_{d12}  \Ccirc \left(   a_{1T} + 2 a_{2T} \vmr  \right) - G_{d21} \MY_2^+ \\
	{\bf S_2^-}	&=	&	G_{d12}  \Ccirc \left(   a_{1T} + 2 a_{2T} \vmr  \right) - G_{d21} \MY_2^- \, .
\end{array}
\right.
	\label{TF_opt2}
\end{align}

\section{Simulation setup}
\label{appendix_simu_setup}

\renewcommand{\toprule}  { \hline\noalign{\smallskip} }
\renewcommand{\midrule} { \noalign{\smallskip}\hline\noalign{\smallskip} }
\renewcommand{\bottomrule} {\noalign{\smallskip}\hline}

\begin{table}[h!]
\centering
	\caption{Electrical parameters of the first 100GHz \HFI{} channel
	}
\begin{tabular}{	 l l l l }
    	\toprule
		 Name		&	Value				&	Unit 	&	Description \\
	\midrule
		$f_{mod}$		&	$90.18685$ 			&	Hz	&	modulation frequency\\
		$vtri$		&	$0.62399$			&	V	&	triangle wave amplitude\\
		$vsqu$		&	$0.17796$			&	V	&	square wave amplitude\\
		$C_0$     		&    	$4.886 \times 10^{-12}$ 	&	F	&	bias capacitance 0	\\
		$C_1$     		&   	$4.711 \times 10^{-12}$ 	&	F  	&	bias capacitance 1	\\
		$C_s$		&	$148.8 \times 10^{-12}$ 	&	F	&	stray capacitance	\\
		$\Ravg$		&	$11.0 \times 10^{6}$		&	$\Omega$ &	bolometer average impedance \\
	\bottomrule
\end{tabular} 
\label{table_elec_params}
\end{table}

\begin{table*}[h!]
\center
\caption{Thermal Parameters  of the first 100GHz \HFI{} channel}
\begin{tabular}{	 L l l l}
    	\toprule
		 Name		&	Value			&	Unit	&	Description \\
	\midrule
		C_0		&	$22.47$ pF			&	J/K	&	heat capacity coefficient\\
		\gamma	&	$1.91$				&	Adimentional	& heat capacity temperature exponent\\
		R_G		&	$57.46$				&	$\Omega$	& bolometer impedance a reference temperature\\
		T_G		&	$11.18$				&	K			& reference temperature for bolometer impedance \\
		\beta		&	$1.3$				&	Adimentional 	&  thermal conductance temperature exponent \\
		G_{s0}	&	$4.533 \times 10^{-11}$	&	W/K			& static thermal conductance \\
		T_{ref}	&	$0.1$				&	K			& thermal conductance reference temperature \\
		T_0		&	$0.1$				&	K			& heat sink temperature \\
		\Tavg	&						&	K			& bolometer average temperature \\
		\Pavg	&	$4.5484 \times 10^{-13}$	&	W			& optical power average \\
	\bottomrule
\end{tabular} 
\label{table_therm_params}

\end{table*}

\section{Numerical results}
\label{sect_num_values}
{
\begin{table}[H]
\center
\caption{Relative signal amplitude for steady state signals and optical excitation signals. 
Values are obtained from run2-run1 simulation signals. }
\begin{tabular}{l l l}
\toprule
	{\bf X } &  {\bf	$\sigma(\overline{X})/\langle X \rangle$ 	}   &   {\bf $\sigma(\widetilde{X})/\sigma(\overline{X})$ } \cr
\midrule
	V     &    \,		   &   3.5e-06 \\
	T     &   5.5e-04    &   1.5e-03 \\
	R    &   3.3e-03    &   1.5e-03 \\
\bottomrule
\end{tabular}
\label{table_numval}
\end{table}

}

\end{appendices}




\bibliography{bibtex}


\end{document}